\documentclass[intlimits,twoside,a4paper]{article}

\usepackage{amsmath,amssymb}
\usepackage{graphicx}

\usepackage[T2A]{fontenc}
\usepackage[cp1251]{inputenc}
\usepackage{xcolor}

\usepackage[eqsecnum]{cmpj2}
\issue{2016}{19}{2}{23603}
\doinumber{10.5488/CMP.19.23603}

\title[Two- and three-phase equilibria]%
{Two- and three-phase equilibria in polydisperse Yukawa hard-sphere mixture.
High temperature and mean spherical approximations}
\author[T.V. Hvozd, Y.V. Kalyuzhnyi]{T.V. Hvozd, Y.V. Kalyuzhnyi}
\address{Institute for Condensed Matter Physics of the National Academy of Sciences of Ukraine, \\ 1 Svientsitskii St.,  79011 Lviv, Ukraine
}

\date{Received October 16, 2015, in final form December 31, 2015}
\authorcopyright{T.V. Hvozd, Y.V. Kalyuzhnyi, 2016}

\begin{document}

\maketitle

\begin{abstract}

Phase behavior of the Yukawa hard-sphere polydisperse mixture with high degree of polydispersity is studied
using high temperature approximation (HTA) and mean spherical approximation (MSA). We have extended and applied
the scheme developed to calculate the phase diagrams of polydisperse mixtures described by the truncatable
free energy models, i.e., the  models with Helmholtz free energy defined by the finite number of the
moments of the species distribution function. At high degree of polydispersity, several new features in the
topology of the two-phase diagram have been observed: the cloud and shadow curves
intersect twice and each of them forms a closed loop of the ellipsoidal-like shape with the liquid and gas
branches of the cloud curve almost coinciding. Approaching a certain limiting value of the polydispersity index,
the cloud and shadow curves shrink and disappear. Beyond this limiting value, polydispersity induces the appearance of the three-phase equilibrium at lower temperatures.
We present and analyze
corresponding phase diagrams together with distribution functions of three coexisting phases.
In general, good agreement was observed between predictions of the two different theoretical methods, i.e., HTA and MSA. Our results confirm qualitative predictions for the three-phase coexistence obtained earlier within the
framework of the van der Waals approach.
\keywords polydispersity, phase coexistence, colloidal systems, Yukawa potential
\pacs 64.75.-g, 82.70.Dd
\end{abstract}

\section{Introduction}

Much of the efforts have been recently focused on the investigation of the effects of polydispersity
on the phase behavior of the colloidal and polymeric fluids.
Polydispersity is an intrinsic property of the colloidal and polymeric systems, since each of the particle
is unique in its shape, size, charge, chain length, etc.
In theoretical description, polydisperse fluids are usually treated as a system consisting of an infinite
number of species with a species index $\xi$ (polydispersity attribute) changing continuously
\cite{Salacuse1982}. Such a system is
characterized by the distribution function of species $f(\xi)$, with $f(\xi)\rd\xi$ being the fraction of
the particles with the value of polydispersity attribute $\xi$ in the range $[\xi,\xi+\rd\xi]$.
Several different theoretical methods for the phase behavior description of polydisperse fluids
have been developed and used recently
\cite{Bellier2000,Leroch1999,Bellier2001,Xu2002,Sollich2002,Wilding2002,Speranza2002,Kalyuzhnyi2003,Wilding2004,Kalyuzhnyi2004,Fasolo2004,Paricaud2004,Fasolo2005,Kalyuzhnyi2005a,Kalyuzhnyi2005b,Wilding2005,Wilding2006,Wilding2010,Wilding2011,Kalyuzhnyi2006,Kalyuzhnyi2007a,Kalyuzhnyi2007b,Kalyuzhnyi2008a,Kalyuzhnyi2008b,Hvozd2015}.
Majority of the earlier studies are based on the van der Waals (vdW)
\cite{Bellier2000,Bellier2001,Xu2002,Sollich2002,Wilding2004,Wilding2005,Wilding2006} or Onsager
\cite{Sollich2002,Speranza2002,Fasolo2004,Fasolo2005} levels of description.
Although such a description appears to be relatively successful in giving qualitative predictions for
the phase behavior of polydisperse mixtures, it is not capable of  providing quantitatively accurate
results.
The major challenge for a theoretical description of the phase behavior of polydisperse mixtures is due to
the functional dependence of Helmholtz free energy of the system $A$ on the distribution function $f(\xi)$.
As a result, phase equilibrium conditions are formulated in terms of the set of an infinite number of equations,
e.g. chemical potential $\mu(\xi)$ for each value of the continuous species index $\xi$ should be the same in each of the coexisting phases. Solution of such a set of equations for Helmholtz free energy of arbitrary
form is next to impossible. However, there is a large number of systems that can be successfully described
using the so-called truncatable free energy (TFE) models \cite{Sollich2002}.
These are models with Helmholtz free energy
represented in terms of a finite number of generalized moments of the distribution function $f(\xi)$.
The latter feature of the TFE models enables one to formulate the phase equilibrium conditions in terms of
a finite set of equations for these moments. Using the scheme developed by Xu et al. \cite{Bellier2000}, the phase
behavior of a number of polydisperse colloidal and polymeric mixtures have been studied using modern methods
of the liquid state theory, which are capable of providing quantitative predictions. These include the use of the mean spherical approximation (MSA)
\cite{Kalyuzhnyi2003,Kalyuzhnyi2004,Kalyuzhnyi2005a,Kalyuzhnyi2005b},
high temperature approximation (HTA)
\cite{Kalyuzhnyi2006,Kalyuzhnyi2007b}
and
thermodynamic perturbation theory
\cite{Paricaud2004,Kalyuzhnyi2007a,Kalyuzhnyi2008a,Kalyuzhnyi2008b,Hvozd2015}. In these papers, two-phase  equilibrium in the systems with low and intermediate degree of polydispersity was studied.

Recently, three-phase equilibrium in polydisperse colloidal mixture with high polydispersity was described
using the scheme based on the vdW approach \cite{Bellier2001}.
In our paper, we propose a corresponding extension of
the scheme used in \cite{Bellier2001} within the framework of the HTA and MSA. We use the scheme developed to study two- and three-phase equilibria in the colloidal mixture with high degree of polydispersity. The paper is organized as follows: in the next section we describe the model and in section~3 we
present the HTA and MSA expressions for thermodynamic properties of the model at hand and discuss the scheme to be used in two- and three-phase diagram calculations. In section~4 we present and discuss the results of these calculations. Our conclusions are collected in section~5. We also present two appendices, in which expressions for the parameters appearing in the expressions for thermodynamic properties, are presented.

\section{The model}
\label{sec:2}

We consider the fluid with interparticle interaction represented by the following multi-Yukawa hard-sphere potential
\begin{equation}
\label{MHC1}
U_\text{HSY}(\xi,\xi';r)=\left\{\begin{array}{ll}
                       \infty, \qquad r\leqslant \sigma(\xi,\xi'), \\[1ex]
                       \displaystyle -\frac{\epsilon_0}{r}\frac{A(\xi)A(\xi')}{z}\re^{-z[r-\sigma(\xi,\xi')]}, \qquad r>\sigma(\xi,\xi'),
                      \end{array}
                   \right.
\end{equation}
where $\xi$ is the polydispersity attribute, i.e., continuous version of the species index,
$\sigma(\xi)$ is the hard-sphere diameter of the particle of species $\xi$, $\sigma(\xi,\xi')=[\sigma
(\xi)+\sigma(\xi')]/2$, $z$ and $\epsilon_0$ are the screening length and the interaction
strength of the Yukawa potential, respectively. The fluid is characterized by the
temperature $T$ [or $\beta=(k_\text{B}T)^{-1}$, where $k_\text{B}$ is the Boltzmann's constant], the total number-
density $\rho$, and by the species distribution function  $f(\xi)$ [$\int f(\xi)\rd\xi=1$].

\section{Theory}
\label{sec:3}

Thermodynamic properties of the model at hand will be described here using HTA and MSA.
Both theories have been used earlier \cite{Kalyuzhnyi2003,Kalyuzhnyi2006} to study the phase
behavior of polydisperse colloidal mixtures with the pair potential
of the type specified above (\ref{MHC1}). Therefore, we will
present here only the final expressions for the quantities needed in our phase equilibrium calculations.
For more details of the HTA and MSA approaches we refer the reader to the original publications.

\subsection{High temperature approximation}

According to the HTA, Helmholtz free energy of
the system $A$ can be written as a sum of two terms: free energy of the reference system
($A_\text{ref}$) and the perturbation term describing the contribution to the free energy due to Yukawa
potential ($A_{1}$):
\begin{equation}
\label{FRE total}
 A=A_\text{ref}+A_{1}=A_\text{HS}+A_{1}.
\end{equation}
Here, $A_\text{ref}=A_\text{HS}$, where $A_\text{HS}$ is the free energy of the hard-sphere fluid, and for $A_{1}$ we have:
\begin{eqnarray}
\label{FRE1a}
\frac{\beta A_{1}}{V}=-\frac{2\pi\beta\epsilon_{0}}{z}\int \rd\xi\int \rd\xi'\rho{(\xi)}\rho{(\xi')}
A(\xi)A(\xi')\widetilde{G}^\text{(HS)}(\xi,\xi';z),
\end{eqnarray}
where $\widetilde{G}^\text{(HS)}(\xi,\xi';z)$ is the Laplace transform of the hard-sphere radial distribution function
\begin{equation}
\label{Gtrdef}
\widetilde{G}^\text{(HS)}(\xi,\xi';z)=\re^{z\sigma(\xi,\xi')}\int_{0}^{\infty} \rd r \, r \, \re^{-zr}g_\text{(HS)}(\xi,\xi';r).
\end{equation}
Here, we use Percus-Yevick approximation for the hard-sphere radial distribution function,
since analytical expressions for its Laplace transform is known.  All the rest of thermodynamical quantities
can be obtained using the expression for Helmholtz free energy (\ref{FRE total}) and standard thermodynamic
relations, e.g., differentiating $A$ with respect to the density we get the expression for the chemical potential:
\begin{equation}
\label{chem total}
\beta\mu(\xi)=\frac{\delta}{\delta\rho(\xi)}\bigg(\frac{\beta A}{V}\bigg),
\end{equation}
and the expression for the pressure $P$ of the system can be calculated invoking the following general
relation:
\begin{eqnarray}
\label{pressure}
\beta P=\beta \int \rd\xi\rho(\xi)\mu(\xi)-\frac{\beta A}{V}.
\end{eqnarray}
In the above expressions, $A_\text{HS}$ and $\mu(\xi)^\text{(HS)}$ are calculated using the corresponding Mansoori et al. expressions \cite{Mansoori}.

Within the framework of the HTA approach, the model in question belongs to the class of  `truncatable free energy (TFE) models', i.e., the models with thermodynamic properties (Helmholtz free energy,
chemical potential, pressure) defined by a finite number of  generalized moments. In this study, we have the following moments:
\begin{equation}
\label{Mlpol}
m_{l,0}=\int \rd\xi\rho(\xi)m_{l,0}(\xi)f(\xi), \qquad   m_{l,0}(\xi)=\sigma^l(\xi),
\end{equation}
\begin{equation}
\label{Mlnpol}
m_{l,\varphi}=\int \rd\xi \rho(\xi)m_{l,\varphi}(\xi)f(\xi), \qquad m_{l,\varphi}(\xi)=\sigma^l(\xi)\varphi(z,\sigma(\xi)), \qquad
\varphi(z,\sigma(\xi))=\frac{1}{z^2}\left[1-z\sigma(\xi)-\re^{-z\sigma(\xi)}\right],
\end{equation}
\begin{equation}
\label{Mlnmpol}
m_{l,A}=\int \rd\xi \rho(\xi)m_{l,A}(\xi)f(\xi), \qquad
m_{l,A}(\xi)=\sigma^l(\xi)A(\xi).
\end{equation}

Closed form analytical expressions for thermodynamic properties (Helmholtz free energy, chemical potential, pressure) in terms of generalized moments (\ref{Mlpol})--(\ref{Mlnmpol}) are presented in the Appendix.

\subsection{Mean spherical approximation}

MSA theory consists of the Ornstein-Zernike (OZ) equation
\begin{eqnarray}
\label{OZeq}
h(r_{12};\xi_1,\xi_2)=c(r_{12};\xi_1,\xi_2)+\rho\int_0^\infty\;\rd\xi_3f(\xi_3)\int \rd{\bf r}_3
c(r_{13};\xi_1,\xi_3)h(r_{32};\xi_3,\xi_2)
\end{eqnarray}
and the following closure relations:
\begin{equation}
\left\{\begin{array}{lll}
c(r_{12};\xi_1,\xi_2)=
& \displaystyle {\beta\epsilon_0\over r_{12}}
{A(\xi) A(\xi')\over z}\re^{-z[r_{12}-\sigma(\xi,\xi')]} ,&\qquad  r_{12}>\sigma(\xi,\xi'),
\\[2ex]
h(r_{12};\xi_1,\xi_2)=&-1,& \qquad
r_{12}\leqslant\sigma(\xi,\xi').
\end{array} \right.
\label{pair}
\end{equation}

For the multicomponent version of the model, the solution of this set of equations reduces to the solution of one
nonlinear algebraic equation for the scaling parameter $\Gamma$ \cite{Ginoza1998}. As a result, all thermodynamic
properties of the model can be expressed in terms of this scaling parameter \cite{Ginoza1998}. More recently, these
expressions were presented in the form suitable to be used in the phase behavior calculations for the polydisperse version of the model at hand \cite{Kalyuzhnyi2003}. Following \cite{Kalyuzhnyi2003}, we have:
\begin{eqnarray}
\label{freeMSA}
\beta(A-A_\text{HS})=\beta E_\text{Y}+\frac{\Gamma^{2}}{3\pi}\bigg(\Gamma+\frac{3}{2}z\bigg),
\end{eqnarray}
$E_\text{Y}$ and $\Gamma$ are determined as follows:
\begin{eqnarray}
\label{E_Y}
\beta E_\text{Y}=K\Bigg\{\rho\Gamma\int_{0}^{\infty}\rd\xi f(\xi)A(\xi)\lambda(\xi)+\frac{\pi}{2\Delta}\frac{[\lambda_{1}]^2}{1+\phi_{1}}+\Delta_N(\lambda_{0}+\lambda_{1}E_{N}) \Bigg\},
\end{eqnarray}
where $K=-\beta\epsilon_{0}$, $\Delta=1-\eta$, $\eta$ is the packing fraction and
\begin{eqnarray}
\label{E_N}
E_{N}=\frac{z}{2}+\Gamma-\frac{\pi}{2\Delta}\frac{\eta_{1}}{1+\phi_{1}},
\end{eqnarray}
\begin{eqnarray}
\label{pressureMSA}
\beta(P-P_\text{HS})=-\frac{\Gamma^{2}}{3\pi}\bigg(\Gamma+\frac{3}{2}z\bigg)+\frac{\pi K}{2\Delta^2}P_N\bigg(P_N+\frac{2z\Delta}{\pi}\Delta_N\bigg)
\end{eqnarray}
with
\begin{eqnarray}
\label{PN}
P_{N}=\frac{\lambda_{1}-\eta_{1}\Delta_N}{1+\phi_{1}}-\frac{z\Delta}{\pi}\Delta_N.
\end{eqnarray}
Yukawa contribution to the chemical potential:
\begin{eqnarray}
\label{chemMSA}
\frac{\rho\beta}{K}\mu_{YU}^\text{(ex)}(\xi)=\rho\lambda(\xi)[A(\xi)\Gamma+\Delta_N(1+\sigma E_N)]+\frac{\delta\{\Delta_N\}}{\delta\{f(\xi)\}}(\lambda_{0}+\lambda_{1}E_N)+
\frac{\pi\rho}{2\Delta}\frac{\sigma(\xi)\lambda_{1}}{1+\phi_{1}}\nonumber\\
\times\left\{(\lambda_{1}-\Delta_N\eta_{1})\left[\frac{\pi}{6\Delta}\sigma^2(\xi)-\phi(\xi)\right]
\frac{\lambda_{1}-\Delta_N\eta_{1}}{1+\phi_{1}}+2\lambda(\xi)-\Delta_N\eta(\xi)\right\}.
\end{eqnarray}
$\Gamma$ satisfies the following nonlinear algebraic equation:
\begin{eqnarray}
\label{GAMMA}
\Gamma^2+z\Gamma=-\pi K D,
\end{eqnarray}
with
\begin{eqnarray}
\label{DDD}
D=\rho\int_{0}^{\infty}\rd\xi f(\xi)[X(\xi)]^2,
\end{eqnarray}
and
\begin{eqnarray}
\label{XXX}
X(\xi)=\lambda \sigma(\xi)-\frac{\lambda_{1}\phi(\xi)}{1+\phi_{1}}-\Delta_N\bigg(\eta(\xi)-\frac{\eta_{1}\phi(\xi)}{1+\phi_{1}}\bigg),
\end{eqnarray}
where $\Delta_N$, $\lambda(\xi)$, $\lambda_{0}$, $\lambda_{1}$, $\phi(\xi)$, $\phi_{0}$, $\phi_{1}$, $\eta(\xi)$, $\eta_{0}$, $\eta_{1}$ are presented in the Appendix.

\subsection{Phase equilibrium conditions}

We assume that at a certain temperature $T$, the system, which is characterized by the parent density $\rho^{(0)}$ and parent-phase distribution function $f^{(0)}(\xi)$ separates into $q$ daughter phases with the densities $\rho^{(1)}$, $\rho^{(2)}$, \ldots , $\rho^{(q)}$, and $q$ daughter distributions $f^{(1)}(\xi)$, $f^{(2)}(\xi)$, \ldots , $f^{(q)}(\xi)$.
Phase equilibrium conditions imply the equality of the pressure $P$ and chemical potential $\mu(\xi)$
in each of $q$ phases, i.e.,
\begin{eqnarray}
\label{equalpressure}
P^{(1)}(T,[f^{(1)}(\xi)])=P^{(2)}(T,[f^{(2)}(\xi)])= \ldots =P^{(q)}(T,[f^{(q)}(\xi)])
\end{eqnarray}
and
\begin{eqnarray}
\label{equalchem}
\mu^{(1)}(\xi,T,[f^{(1)}(\xi)])=\mu^{(2)}(\xi,T,[f^{(2)}(\xi)])= \ldots =\mu^{(q)}(\xi,T,[f^{(q)}(\xi)]).
\end{eqnarray}
Here, $[f(\xi)]$ denote the functional dependence of the corresponding quantity on the distribution function
$f(\xi)$. The phase separation is constrained by the conservation of the total number of the particles
of each species $\xi$ and by the conservation of the total volume occupied by the parent phase, i.e.,
\begin{eqnarray}
\label{conspart}
f^{(0)}(\xi)=\sum_{\alpha}^{q}f^{(\alpha)}(\xi)x^{(\alpha)}
\end{eqnarray}
and
\begin{eqnarray}
\label{consvol}
\upsilon_{0}=\sum_{\alpha}^{q}\upsilon^{(\alpha)}x^{(\alpha)},
\end{eqnarray}
where $x^{(\alpha)}=N^{(\alpha)}/N^{(0)}$ is the ratio of the total number of the particles in the phase
$\alpha$ to the total number of particles in the parent phase, $\upsilon^{(\alpha)}=1/\rho^{(\alpha)}$, $(\alpha=1,2,3, \ldots , q)$. Finally, the normalization of the distribution function $f^{(\alpha)}(\xi)$
\begin{eqnarray}
\label{norm}
\int f^{(\alpha)}(\xi)\rd\xi=1
\end{eqnarray}
in equation (\ref{conspart}) implies the conservation of the total number of the particles:
\begin{eqnarray}
\label{consparticl}
1=\sum_{\alpha}^{q}x^{(\alpha)}.
\end{eqnarray}

Solution of the above set of equations (\ref{equalpressure})--(\ref{consparticl}) gives the number densities
of the coexisting phases $\rho^{(q)}$ and their distribution functions $f^{(q)}(\xi)$. However, due to the
functional dependence of Helmholtz free energy of the system on the distribution function $f(\xi)$, solution of
this set of equations without further simplifications is next to impossible. Substantial simplification occurs
for the TFE models: for this class of the models, the set of functional equations
(\ref{equalpressure})--(\ref{consparticl}) can be written as a set of equations for the generalized moments of
the distribution function $f(\xi)$ \cite{Sollich2002,Bellier2000,Bellier2001,Xu2002}. In this study, we will follow the scheme developed
by Xu and Baus \cite{Bellier2000,Bellier2001,Xu2002} and present a  set of equations (\ref{equalpressure})--(\ref{consparticl})
for two- and three-phase equilibria in terms of  generalized moments (\ref{Mlpol})--(\ref{Mlnmpol}).
{For polydisperse Yukawa hard-sphere mixture with size and interaction
energy polydispersity,}
within the framework of the HTA thermodynamics of the system depends on the set of {9} generalized moments
{(\ref{Mlpol})--(\ref{Mlnmpol})} and within the framework of MSA we have 10 generalized moments {(\ref{phin})--(\ref{phin1})} and one MSA scaling parameter
$\Gamma$ (see Appendices I and II, respectively). In what follows we will conventionally denote these moments
as $m_k$ and use the following definition:
\begin{eqnarray}
\label{momdef}
m_k=\rho\int m_k(\xi)f(\xi)\rd \xi.
\end{eqnarray}
Note, that this set includes also the total number density of the system, i.e.,
$m_0(\xi)=1$ and $m_0=\rho$.

\subsection{Two-phase coexistence}

In the case of the two-phase equilibrium ($\alpha=1, 2$), conditions (\ref{equalpressure})--(\ref{consparticl}) lead to the following set of equations in terms of the generalized moments $m_k$:
\begin{eqnarray}
\label{equalpressure1}
P^{(1)}(T,\{m_k^{(1)}\})=P^{(2)}(T,\{m_{k}^{(2)}\}),
\end{eqnarray}
\begin{eqnarray}
\label{momfin1}
m_{k}^{(1)}=m_{0}^{(1)}\int m_{k}^{(1)}(\xi)f^{(0)}(\xi)H(\xi,T,m_{0}^{(2)},\{m^{(1)}\}\{m^{(0)}\})\rd\xi, \qquad k\neq0 ,
\end{eqnarray}
\begin{eqnarray}
\label{momfin2}
m_{k}^{(2)}=m_{0}^{(2)}\int m_{k}^{(2)}(\xi)f^{(0)}(\xi)\left[\frac{v_2-v_1}{v_0-v_1}+
\frac{v_2-v_0}{v_1-v_0}H(\xi,T,m_{0}^{(2)},\{m^{(1)}\}\{m^{(0)}\})\right]
\rd\xi, \qquad k\neq0 ,
\end{eqnarray}
\begin{eqnarray}
\label{norm1}
\int f^{(\alpha)}(\xi)\rd\xi=1, \qquad  \text{for} \qquad  \alpha=1 \quad \text{or} \quad \alpha=2,
\end{eqnarray}
with $k=1,\ldots, {8}$ for HTA and $k=1,\ldots , {9}$ for MSA. In the latter case, this set of equations is
supplemented by the following two equations for the scaling parameters $\Gamma^{(1)}$ and $\Gamma^{(2)}$
\begin{equation}
\label{Gamma12}
\Gamma^{(\alpha)}=-\frac{\pi KD^{(\alpha)}}{z+\Gamma^{(\alpha)}}\,, \qquad \alpha=1,2.
\end{equation}
Here, $\left\{m^{(q)}\right\}$ denote the set of all moments of the phase $q$,
\begin{eqnarray}
\label{fdistr1}
f^{(1)}(\xi)=f^{(0)}(\xi)H(\xi,T,m_{0}^{(2)},\{m^{(1)}\}\{m^{(0)}\}),
\end{eqnarray}
\begin{eqnarray}
\label{fdistr2}
f^{(2)}(\xi)=\frac{v_2-v_1}{v_0-v_1}f^{(0)}(\xi)+\frac{v_2-v_0}{v_1-v_0}f^{(1)}(\xi),
\end{eqnarray}
\begin{eqnarray}
\label{Hxi}
&&H(\xi,T,m_{0}^{(2)},\{m^{(1)}\}\{m^{(0)}\})=\nonumber\\
&&=\frac{(\rho^{(1)}-\rho^{(2)})
A_{12}(\xi,T,m_{0}^{(2)},\{m^{(1)}\}\{m^{(0)}\})}
{(\rho^{(1)}\rho^{(2)}/\rho^{(0)}-\rho^{(2)})+(\rho^{(1)}-\rho^{(1)}\rho^{(2)}/\rho^{(0)})
A_{12}(\xi,T,m_{0}^{(2)},\{m^{(1)}\}\{m^{(0)}\})},
\end{eqnarray}
\begin{eqnarray}
\label{Axi}
A_{12}(\xi,T,\{m^{(1)}\}\{m^{(2)}\})=\frac{\rho^{(2)}}{\rho^{(1)}}
\exp\left[\mu_\text{ex}^{(2)}(\xi,T,\{m^{(2)}\})-\mu_\text{ex}^{(1)}(\xi,T,\{m^{(1)}\})\right] ,
\end{eqnarray}
and $\mu_\text{ex}^{(q)}$ is the value of the excess chemical potential (to its ideal gas value) in the phase $q$.
Solution of this two-phase problem [equations (\ref{equalpressure1}), (\ref{norm1}) and (\ref{momfin1}), (\ref{momfin2})], for a given temperature $T$, the density of the parent phase $\rho^{(0)}$, and the parent species
distribution function $f^{(0)}(\xi)$ gives the coexisting densities $\rho^{(\alpha)}$ of the two daughter phases and corresponding species distribution functions $f^{(\alpha)}(\xi)$, $\alpha=1,2$.

The coexisting densities for different temperatures give binodals, which are terminated  at a temperature for which the density of one of the phases is equal to the density $\rho^{(0)}$ of the parent phase. These termination points form the cloud and shadow coexisting curves, which intersect at the critical point, {with} the critical temperature $T_\text{cr}$ and the critical density
$\rho_\text{cr}=\rho^{(1)}=\rho^{(2)}=\rho^{(0)}$.
The cloud and shadow curves can be obtained as a special solution of the
coexisting problem, when the properties of one phase are
equal to the properties of the parent phase: assuming that the
phase $2$ is the cloud phase, i.e., $\rho^{(2)}=\rho^{(0)}$, and following
the above scheme we will end up with the same set of equations, but with $\rho^{(2)}$ and $f^{(2)}(\xi)$ substituted by $\rho^{(0)}$ and $f^{(0)}(\xi)$, respectively.

\subsection{Three-phase coexistence}

In the case of three-phase equilibrium, the phase coexistence conditions  (\ref{equalpressure})--(\ref{consparticl}) are reduced to the following set of equations:
\begin{eqnarray}
\label{equalpressure3ph1}
P^{(1)}(T,\{m_k^{(1)}\})=P^{(2)}(T,\{m_{k}^{(2)}\})=P^{(3)}(T,\{m_{k}^{(3)}\}),
\end{eqnarray}
\begin{eqnarray}
\label{norm3ph1}
\int f^{(1)}(\xi)\rd\xi=1,
\qquad
\int f^{(2)}(\xi)\rd\xi=1,
\end{eqnarray}
\begin{eqnarray}
\label{momfin3ph1}
m_{k}^{(\alpha)}=m_{0}^{(\alpha)}\int m_{k}^{(\alpha)}(\xi)f^{(0)}(\xi)H_{\alpha}(\xi)\rd\xi,\qquad
k\neq0,\qquad \alpha=1,2,3.
\end{eqnarray}
with $k=1,\ldots, {8}$ for HTA and $k=1,\ldots, {9}$ for MSA. In the latter case, this set of equations is
supplemented by the following three equations for the scaling parameters $\Gamma^{(1)}$,  $\Gamma^{(2)}$ and  $\Gamma^{(3)}$
\begin{equation}
\label{Gamma123}
\Gamma^{(\alpha)}=-\frac{\pi KD^{(\alpha)}}{z+\Gamma^{(\alpha)}}\;,\;\;\;\;\;\alpha=1,2,3.
\end{equation}
Here,
\begin{eqnarray}
\label{fdistr3ph1}
f^{(1)}(\xi)=f^{(0)}(\xi)H_1(\xi),\;\;\;\;\;H_1(\xi)=\frac{A_{13}(\xi)}{x_1A_{13}(\xi)+x_2A_{23}(\xi)+x_3},
\end{eqnarray}
\begin{eqnarray}
\label{fdistr3ph2}
f^{(2)}(\xi)=f^{(0)}(\xi)H_2(\xi),\;\;\;\;\;H_2(\xi)=\frac{A_{23}(\xi)}{x_1A_{13}(\xi)+x_2A_{23}(\xi)+x_3},
\end{eqnarray}
\begin{eqnarray}
\label{fdistr3ph3}
f^{(3)}(\xi)=f^{(0)}(\xi)H_3(\xi),\;\;\;\;\;H_3(\xi)=\frac{1}{x_1A_{13}(\xi)+x_2A_{23}(\xi)+x_3},
\end{eqnarray}
\begin{eqnarray}
\label{Axi1}
A_{13}(\xi)=\frac{\rho^{(3)}}{\rho^{(1)}}
\exp\left[\mu_\text{ex}^{(3)}(\xi,T,\{m^{(3)}\})-\mu_\text{ex}^{(1)}(\xi,T,\{m^{(1)}\})\right],
\end{eqnarray}
\begin{eqnarray}
\label{Axi2}
A_{23}(\xi)=\frac{\rho^{(3)}}{\rho^{(2)}}
\exp\left[\mu_\text{ex}^{(3)}(\xi,T,\{m^{(3)}\})-\mu_\text{ex}^{(2)}(\xi,T,\{m^{(2)}\})\right],
\end{eqnarray}
\begin{eqnarray}
\label{fr3ph}
x_1=1-x_3-\frac{v_0-v_1+x_3(v_1-v_3)}{v_2-v_1}, \qquad x_2=\frac{v_0-v_1+x_3(v_1-v_3)}{v_2-v_1}.
\end{eqnarray}

Solution of this set of equations for a given temperature $T$, density of the parent phase $\rho^{(0)}$, and parent species distribution function $f^{(0)}(\xi)$ gives the coexisting densities $\rho^{(\alpha)}$ of the three daughter phases and the corresponding species distribution functions $f^{(\alpha)}(\xi)$, $\alpha=1,2,3$.

\section{Results and discussion}
\label{sec:4}

In this section we present numerical results for the phase behavior of polydisperse Yukawa
hard-sphere mixture.
To complete description of the system, we have chosen log-normal
distribution for the species distribution function $f(\xi)$ of the parent phase, i.e.,
\begin{equation}
\label{distrF}
 f^\text{(LN)}(\xi)=\frac{I}{\sqrt{2\pi \ln I}}\exp\left[-\frac{\ln^{2}(I^{3/2}\xi)}{2\ln I}\right],
\end{equation}
where $I$ is the polydispersity index. {Log-normal distribution is often used to describe polydisperse fluids \cite{Elias,Hunter}.} In the monodisperse limit ($I=1$), this
distribution is represented by the Dirac delta-function, $\delta(\xi-1)$.
On the contrary, when $I$ becomes very large ($I\gg1$), the above distribution becomes very
wide, increasing the contributiion to the phase behavior of the particles with a large value of $\xi$.
All calculations were carried out for the one-Yukawa version of the pair potential (\ref{MHC1}) with $z=1.8\sigma_0$.
We consider polydispersity only in the
strength (amplitude) of the pair potential $A(\xi)$. In this case, we have chosen
$A(\xi)=A_0\xi$ and $\sigma(\xi)=\sigma_0$. Here, $A_0=1$ and $\sigma_0$ is the hard-sphere size for a monodisperse version of the model at $I=1$, which is used as a distance unit.
In what follows, the density $\rho$ and temperature $T$ are presented in reduced units, i.e., $\rho^*=\rho\sigma_0^3$ and $T^*=k_\text{B}T/\epsilon_0$.

\subsection{Two-phase coexistence}
\label{sec:4.1}

\begin{figure}[!t]
\centerline{
\includegraphics[width=0.5\textwidth]{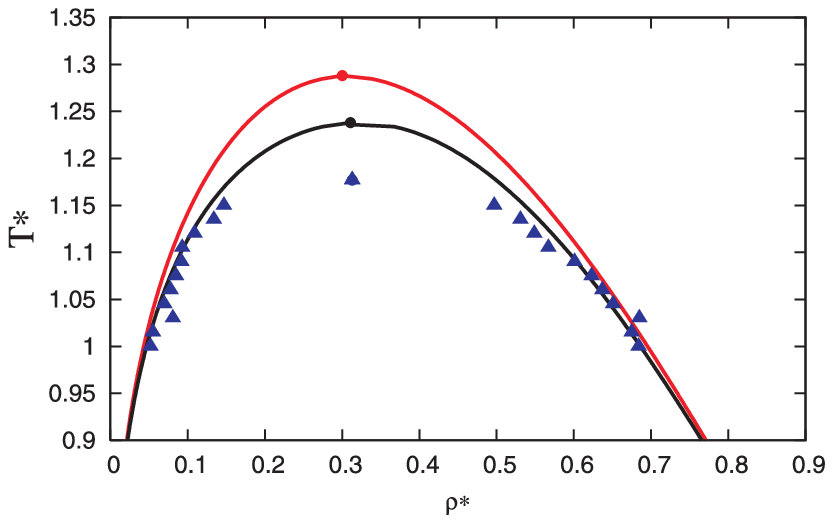}
}
\centerline{
\includegraphics[width=0.5\textwidth]{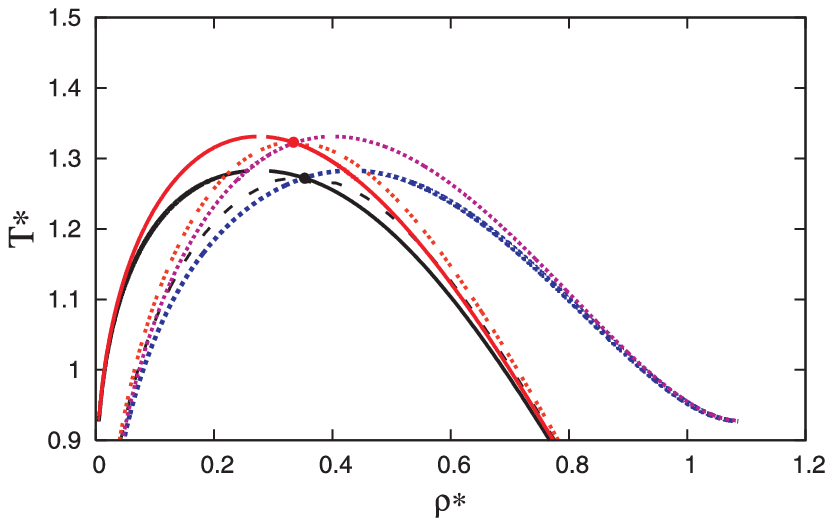}
\includegraphics[width=0.5\textwidth]{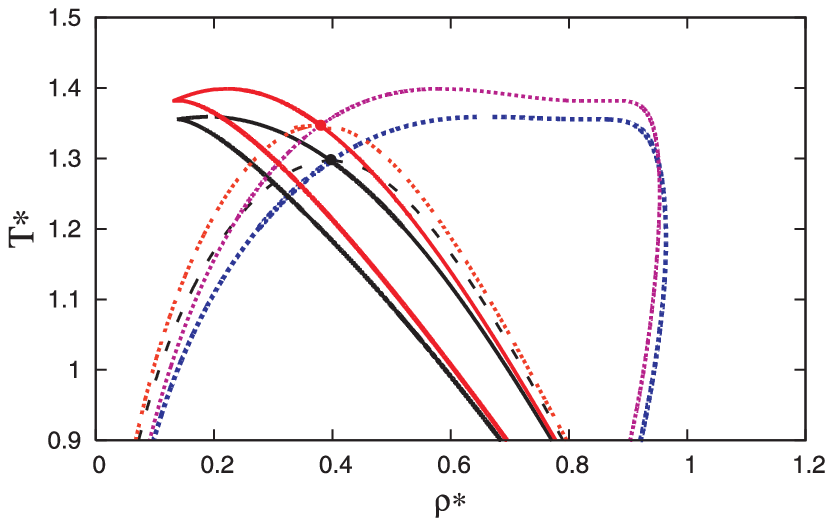}
}
\caption{(Color online) Phase diagrams of the polydisperse Yukawa hard-sphere mixture with amplitude polydispersity only in the $(\rho^*,T^*)$ plane at low ($I=1.01$, the left-hand lower panel) and intermediate ($I=1.02$, the right-hand lower panel) degree of polydispersity,
obtained using HTA (red and purple lines) and MSA (black and blue lines), which includes cloud (solid line) and shadow (dotted line) curves, two critical points and critical binodals (dashed lines). Filled circle denotes the position of the critical points.
In upper panel we demonstrate the liquid-gas phase diagram of monodisperse system ($I=1$), obtained using HTA (full red line), MSA (solid blue line) and computer simulation method (triangles).}
\label{enlnhtalow}
\end{figure}

We calculate the phase diagrams of the polydisperse Yukawa hard-sphere mixture at different values of polydispersity index $I$ using HTA and MSA. In figure~\ref{enlnhtalow} (the upper panel), we show the liquid-gas phase diagram of the
system in the limiting case of $I=1$. In this case, the system is monodisperse, and the corresponding cloud and
shadow curves coincide with binodals, and the critical point is located at their maximum. In addition,
in the same figure we present the phase diagram obtained using the computer simulation method \cite{Pini,Lomba}.
As one would expect, the agreement between computer simulations and theoretical results is reasonably good, with
MSA predictions being a bit more accurate.

\begin{figure}[!t]
\centerline{
\includegraphics[width=0.5\textwidth]{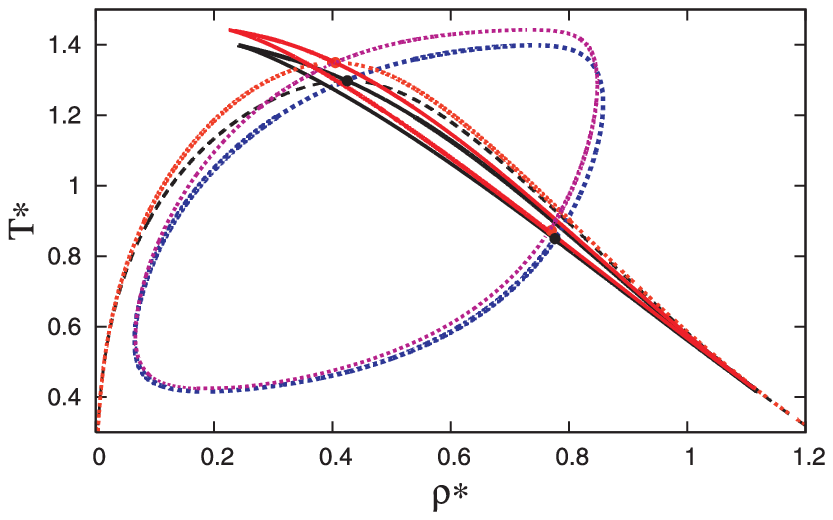}
\includegraphics[width=0.5\textwidth]{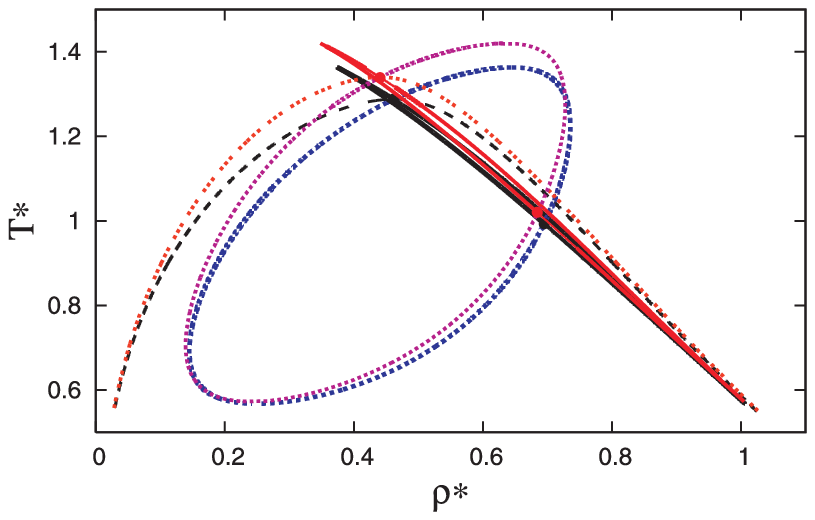}
}
\centerline{
\includegraphics[width=0.5\textwidth]{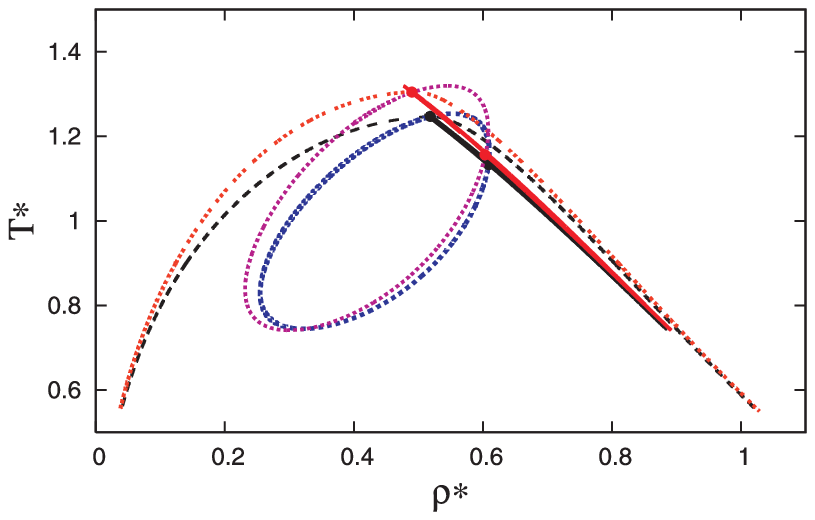}
\includegraphics[width=0.5\textwidth]{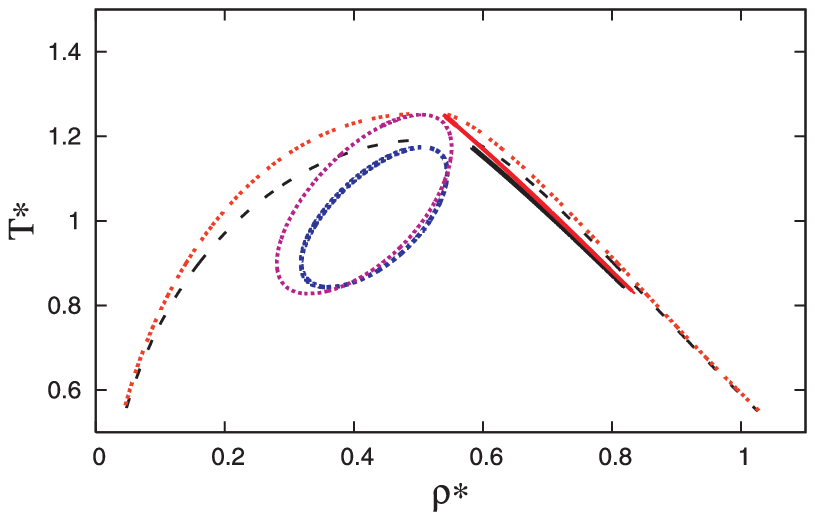}
}
\caption{(Color online) Phase diagrams of the polydisperse Yukawa hard-sphere mixture with amplitude polydispersity only in the $(\rho^*,T^*)$ plane for three different values of the polydispersity index $I$, $I=1.024$ (the left-hand upper panel), $I=1.028$ (the right-hand upper panel), $I=1.031$ (the left-hand lower panel) and $I=1.032$ (the right-hand lower panel), obtained using HTA (red and purple lines) and MSA (black and blue lines), which includes cloud (solid line) and shadow (dotted line) curves, two critical points and critical binodals (dashed lines). Filled circle denotes the position of the critical points.}
\label{enlnhta}
\end{figure}

In figure~\ref{enlnhtalow} (the lower panels), we demonstrate the phase diagram for the
model in question at low and intermediate degree of polidispersity ($I=1.01$ and $I=1.02$). Now, the cloud and shadow curves
are splitted and the position of the binodals depends on the value of the parent phase total number density
$\rho^{(0)}$. Each of the binodals is terminated on either cloud or shadow curves. Corresponding pairs of
the terminal points represent the two phases in equilibrium with infinitesimally small amount of the phase
located on the shadow curve. Critical point is located on the intersection of the cloud and shadow curves
and binodals with critical value of the parent total number density. This type of the phase behavior is
rather standard and have been observed in a number of the previous studies
\cite{Kalyuzhnyi2003,Kalyuzhnyi2006}. However with
a further increase of polydispersity, several novel features appear.
For small polydispersity, the system has only one critical point
\cite{Kalyuzhnyi2003,Kalyuzhnyi2004,Kalyuzhnyi2005a,Kalyuzhnyi2005b,Kalyuzhnyi2006},
which originates from the regular liquid-gas (LG) critical point of the corresponding monodisperse
version of the system. With the polydispersity increase, there appears an additional critical
point, which is induced by the polydispersity (figure~\ref{enlnhta}). This effect on the qualitative
van der Waals level of description has been observed by Bellier-Castella et al.
\cite{Bellier2000,Bellier2001}. The second critical point, which we
denote as polydisperse (P) critical point, is located at larger values of the density and at lower
values of the temperature. With the appearance of the second critical point, the cloud and shadow curves
intersect twice and each of them forms a closed loop of the ellipsoidal-like shape. With an  increase
of polydispersity, liquid and gas branches of the cloud curve approach each other.
{Here, the gas branch of the cloud curve has a negative slope, which
indicates that under these conditions the system may show a reentrant
phase behaviour.}

\begin{table}[!b]
\caption{Parameters of the critical points obtained by HTA and MSA for polydisperse Yukawa hard-sphere mixture with amplitude polydispersity only at three different polydispersity indices $I$. At $I=1.032$, there are no critical points.}
\label{tbl-encrp}
\vspace{2ex}
\begin{center}
\renewcommand{\arraystretch}{0}
\begin{tabular}{|c|c||c|c|c|}
\hline\hline
&&$I=1.024$&$I=1.028$&$I=1.031$\strut\\
\hline
\rule{0pt}{2pt}&&&&\\
\hline
\raisebox{-1.7ex}[0pt][0pt]{HTA $\rho_\text{cr, HTA}^*$}
      & LG& 0.405&  0.44&  0.49\strut\\
\cline{2-5}
      & P& 0.77&  0.684&  0.603\strut\\
\hline
\raisebox{-1.7ex}[0pt][0pt]{HTA $T_\text{cr, HTA}^*$}
      & LG& 1.349&  1.339&  1.304\strut\\
\cline{2-5}
      & P& 0.872&  1.02&  1.157\strut\\
\hline
\raisebox{-1.7ex}[0pt][0pt]{MSA $\rho_\text{cr, MSA}^*$}
      & LG& 0.425& $0.461$&  0.518\strut\\
\cline{2-5}
      & P& 0.776&  0.695&  0.609\strut\\
\hline
\raisebox{-1.7ex}[0pt][0pt]{MSA $T_\text{cr, MSA}^*$}
      & LG& 1.298& $ 1.288$&  1.247\strut\\
\cline{2-5}
      & P& 0.85&  0.993&  1.133\strut\\
\hline\hline
\end{tabular}
\renewcommand{\arraystretch}{1}
\end{center}
\end{table}

In figure~\ref{enlnhta} we present the phase diagrams obtained by HTA and MSA theories, which include two critical points, cloud and shadow curves and critical binodals.
In the left-hand upper panel we show the results for the polydispersity index $I=1.024$. {In table~\ref{tbl-encrp} we present the values of the critical temperature and density obtained by MSA and HTA at three different polydispersity indices $I$.
In the right-hand upper panel (figure~\ref{enlnhta}) we show the results for the polydispersity index $I=1.028$.
With the polydispersity increase ($I=1.031$), both LG and P  critical points move towards each other, (figure~\ref{enlnhta}, left-hand lower panel) and the corresponding cloud and shadow curves shrink, and
for a certain limiting value $(I=1.032)$ of polydispersity, they merge.}
Above this limiting value, there are no critical points (figure~\ref{enlnhta}, lower panel).
The `liquid' and `gas' branches of the cloud curves almost coincide for the larger polydispersity  (figure~\ref{enlnhta}, right-hand lower panel). With a further increase of polydispersity, the cloud and shadow curves shrink and finally disappear.

Both approximations provide quantitatively close predictions for the phase behavior of the model. Polydispersity does not substantially change the relation between MSA and HTA results, i.e., HTA predicts slightly higher critical temperature and slightly lower critical density. We assume that similarly as in monodisperse case (figure~\ref{enlnhtalow}) HTA is less accurate than MSA.
Note that HTA describes correlations between the particles on the level of the reference system (in this study
hard-sphere system) neglecting contribution from the attractive part of the potential to the structure.

\subsection{Three-phase fractionation}

\begin{figure}[!b]
\centerline{
\includegraphics[width=0.5\textwidth]{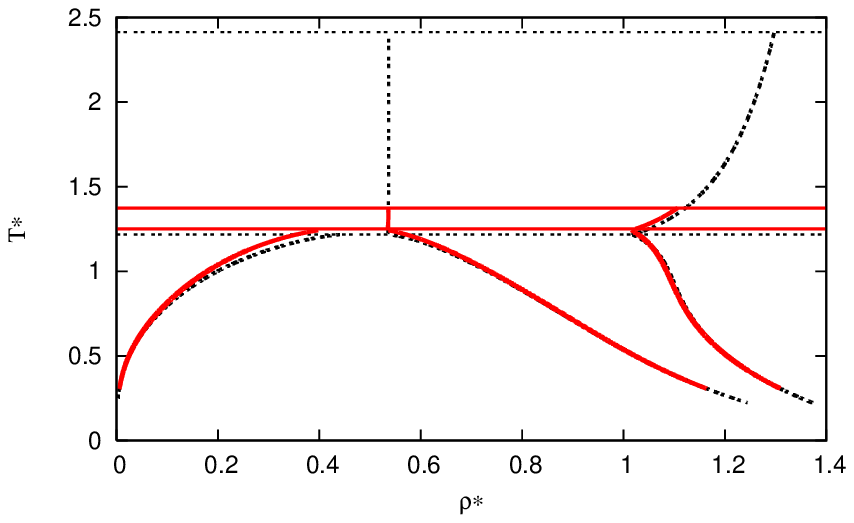}
\includegraphics[width=0.5\textwidth]{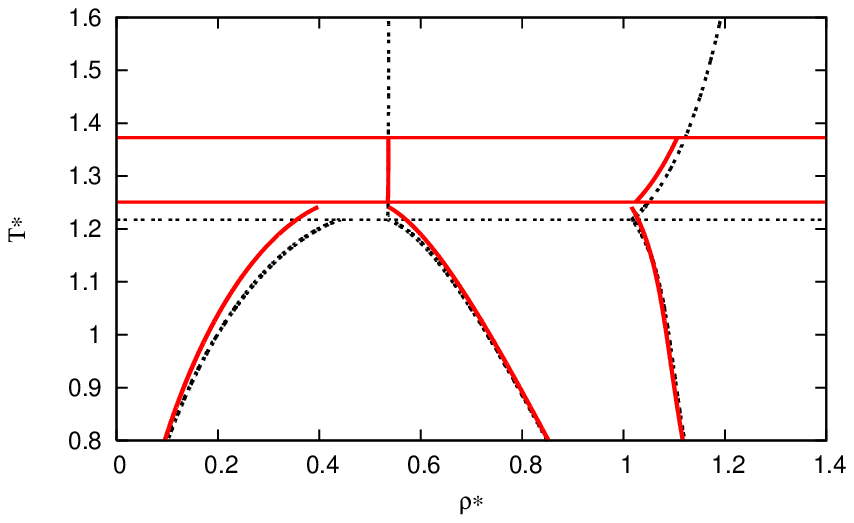}
}
\centerline{
\includegraphics[width=0.5\textwidth]{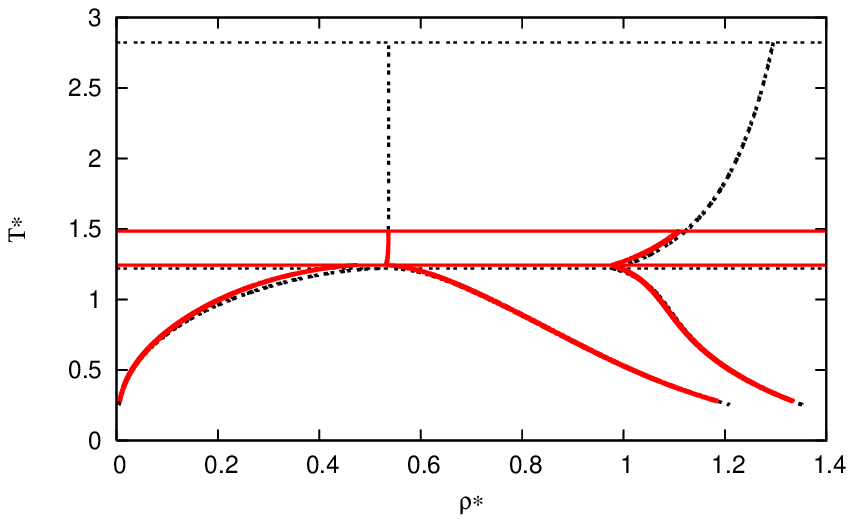}
\includegraphics[width=0.5\textwidth]{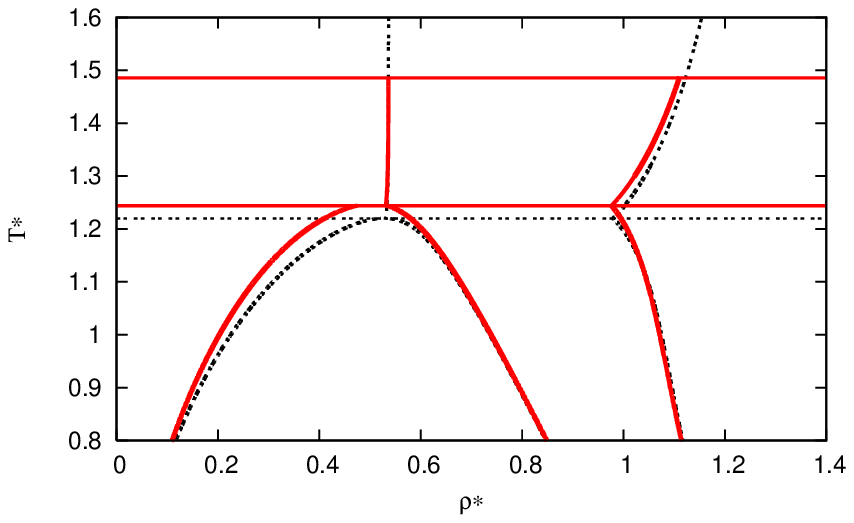}
}
\centerline{
\includegraphics[width=0.5\textwidth]{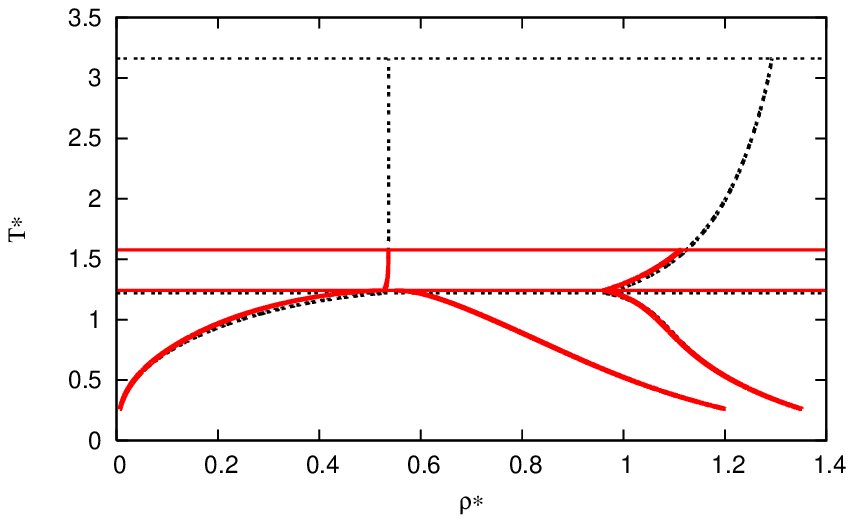}
\includegraphics[width=0.5\textwidth]{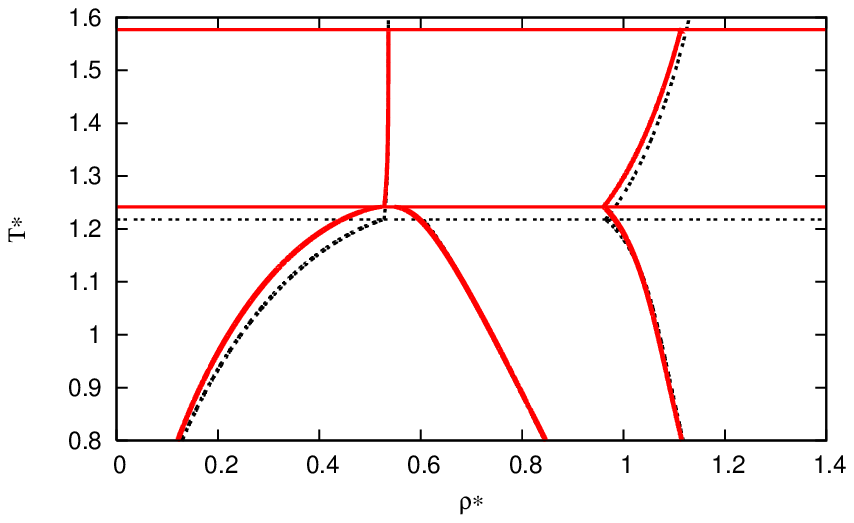}
}
\caption{(Color online) Phase diagrams of polydisperse Yukawa hard-sphere mixture for $I=1.032$ (the upper panel), $I=1.036$ (the intermediate panel), $I=1.039$ (the lower panel) and $\rho_0^*=0.5365$ obtained by HTA (full red line) and MSA (dotted black line).}
\label{msa-hta-full}
\end{figure}

As can be seen in subsection~\ref{sec:4.1}, at relatively large values of polydispersity
index $I$ in the phase behavior of our system, there appear several novel features, which are absent at lower
values of $I$. In particular, at a limiting value of $I$ ($I=1.032$), the critical points disappear.
This appears to be a sign for the possibility of the three-phase equilibria.
For the system with the limiting value of polydispersity $I=1.032$ with the temperature decrease at a
certain temperature, the system is fractionated into three coexisting phases. We have studied this
transition at three different values of polydispersity index $I$, i.e., $I=1.032$, 1.036, 1.039.
In figure~\ref{msa-hta-full} we show the phase diagrams of the system at the set values of $I$ obtained
using  HTA (full line) and MSA (dotted line).
$T_{(I)}^{*(1)}$ denotes the borderline  between the one-phase
and two-phase regions at polydispersity index $I$, $T_{(I)}^{*(2)}$ is the borderline  between the two-phase and three-phase regions. As can be seen in figure~\ref{msa-hta-full}, MSA shifts the HTA $T_{(I),\text{HTA}}^{*(1)}$ borderline to much higher values and $T_{(I),\text{HTA}}^{*(2)}$ to slightly lower values.
{In table~\ref{tbl-temp} we present the results for $T_{(I)}^{*(1)}$ and $T_{(I)}^{*(2)}$ obtained by HTA and MSA theories at three different polydispersity indices $I$. With polydispersity increase, the two-phase region extends to a wider temperature range. This can be seen in figure~\ref{msa-hta-full} and table~\ref{tbl-temp}.}
Note that while in the upper panel the new (third) phase is a low-density phase, in the lower panel it is an intermediate-density phase.

\begin{table}[!t]
\caption{The borderlines obtained by HTA and MSA for polydisperse Yukawa hard-sphere mixture with amplitude polydispersity only at three different polydispersity indices $I$.}
\label{tbl-temp}
\vspace{2ex}
\begin{center}
\renewcommand{\arraystretch}{0}
\begin{tabular}{|c|c||c|c|c|}
\hline\hline
&&$I=1.032$&$I=1.036$&$I=1.039$\strut\\
\hline
\rule{0pt}{2pt}&&&&\\
\hline
\raisebox{-1.7ex}[0pt][0pt]{HTA}
      & $T_{(I)}^{*(1)}$& 1.373&  1.486&  1.577\strut\\
\cline{2-5}
      & $T_{(I)}^{*(2)}$& 1.251&  1.244&  1.242\strut\\
\hline
\raisebox{-1.7ex}[0pt][0pt]{MSA}
      & $T_{(I)}^{*(1)}$& 2.414&  2.823&  3.160\strut\\
\cline{2-5}
      & $T_{(I)}^{*(2)}$& 1.217&  1.220&  1.218\strut\\
\hline\hline
\end{tabular}
\renewcommand{\arraystretch}{1}
\end{center}
\end{table}

As noted above, the polydisperse system can be viewed as a mixture of an infinite number
of components, each characterized by a continuous variable $\xi$. Thus, the Gibbs phase rule allows for a phase fractionation into, say $q$ phases. Let the polydisperse system be fractionated into the ($q-1$)-phase region with defined borderlines  $T_{(I)}^{*(1)}>T_{(I)}^{*(2)}>T_{(I)}^{*(3)}>\ldots>T_{(I)}^{*(q-2)}$. With both, the polydispersity index $I$ and $q$ increase with decreasing $T_{(I)}^{*(q-1)}$, the ($q$)-phase region is fractionated. Note that in the monodisperse limit ($I=1$), our system has only a one-phase (for $T^*>T^*_\text{cr}$) region and a two-phase (for $T^*<T^*_\text{cr}$) region. When $I=1$, all these temperatures ($T_{(I)}^{*(1)},T_{(I)}^{*(2)},T_{(I)}^{*(3)},\ldots,T_{(I)}^{*(q-1)}$) will reduce to $T_{(1)}^{*(1)}=T^*_\text{cr}$.

\begin{figure}[!t]
\centerline{
\includegraphics[width=0.5\textwidth]{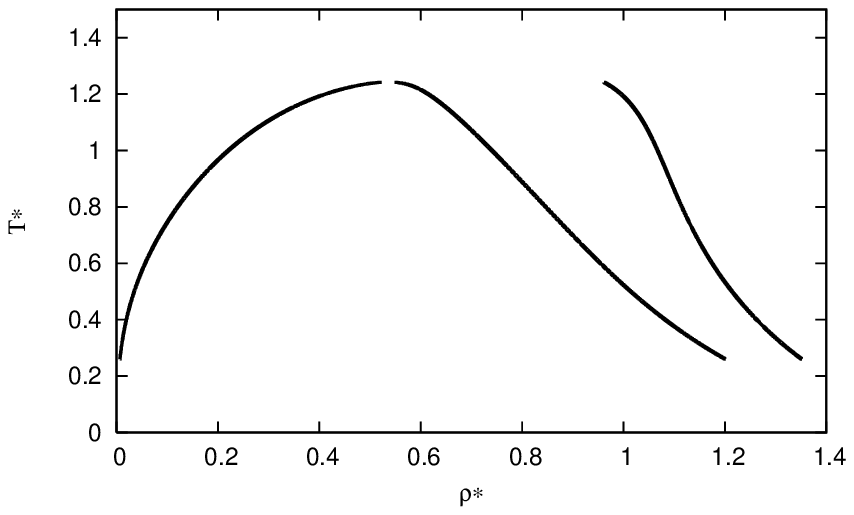}
\includegraphics[width=0.5\textwidth]{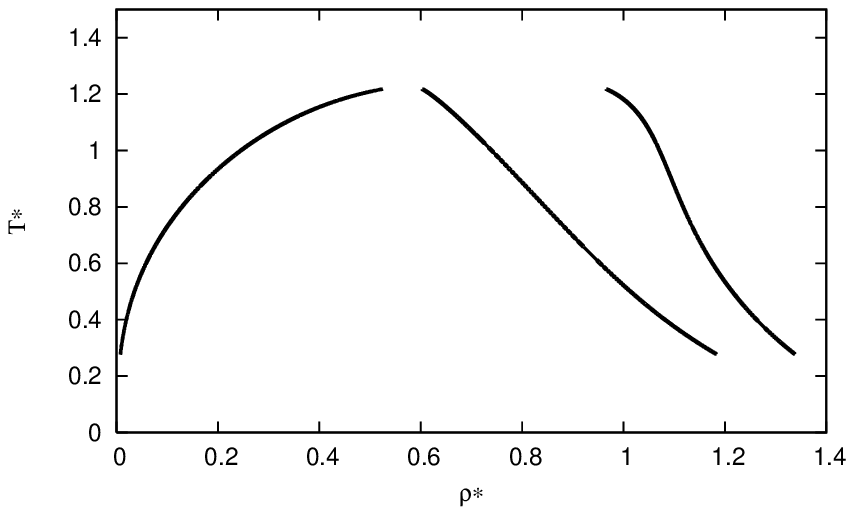}
}
\centerline{
\includegraphics[width=0.5\textwidth]{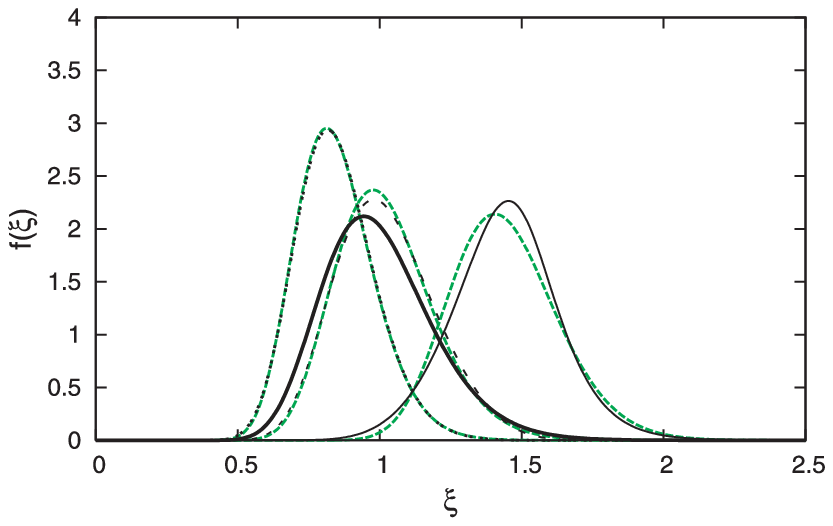}
\includegraphics[width=0.5\textwidth]{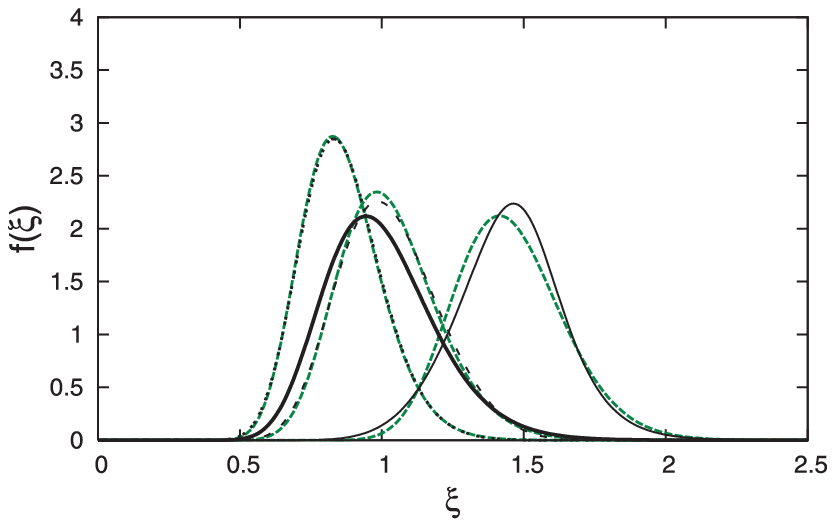}
}
\centerline{
\includegraphics[width=0.5\textwidth]{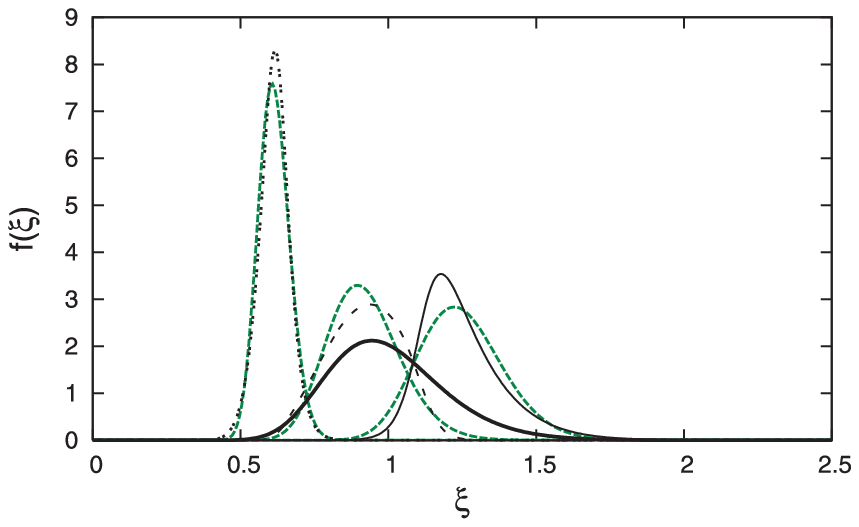}
\includegraphics[width=0.5\textwidth]{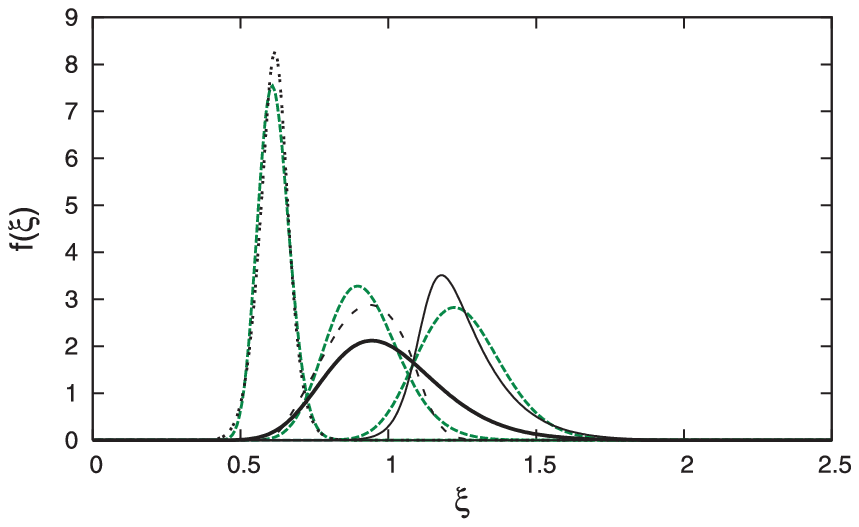}
}
\caption{(Color online) A three-phase coexistence (the upper panels) for $I=1.039$ and $\rho_0^*=0.5365$ is obtained by HTA (the left-hand column) and MSA (the right-hand column), and corresponding coexisting distributions at $T^*=1.05$ (the intermediate panels) and $T^*=0.35$ (the lower panels).
The solid line shows parent $f^{(0)}(\xi)$ distribution, the dotted line corresponds to  a low-density phase, the dashed line displays an intermediate-density phase, and the full line refers to a high-density phase. Green dashed line denotes log-normal distributions obtained from daughter distribution (as defined in the text).}
\label{distr}
\end{figure}

We show how an initial log-normal parent-phase distribution, $f^{(0)}(\xi)$, fractionates into three daughter phases. The degree of fractionation depends on the temperature, and these three daughter distributions,
$f^{(1)}(\xi)$, $f^{(2)}(\xi)$ and $f^{(3)}(\xi)$ change with temperature along the coexistence curve.
In figure~\ref{distr} distribution functions of the polydisperse Yukawa hard-
sphere fluid with amplitude polydispersity at two values of the temperature are reported.
The left-hand column corresponds to the HTA results, and the right-hand column shows the MSA results.
In the upper panels we display the phase diagrams at $I=1.039$ and $\rho_0^*=0.5365$.
We present distribution functions of the coexisting phases on the three-phase coexistence curves
at $T^*=1.05$ (figure~\ref{distr}, intermediate panels) and $T^*=0.35$ (figure~\ref{distr}, lower panels).
Similarly to the case of the two-phase equilibria, the particles with larger values of
$\xi$ (stronger interacting particles) fractionate to the high-density (liquid) phase and the particles with
smaller values of $\xi$ fractionate into the low-density (gas) phase.
With the temperature decrease, all distributions become narrower and their maxima increase.

Finally, we have analyzed the shape of the daughter distribution functions $f^{(1)}(\xi)$, $f^{(2)}(\xi)$, $f^{(3)}(\xi)$. We have investigated how the daughter distribution functions resemble log-normal distributions. To obtain the polydispersity index, $I=m_2^{(\alpha)}/(m_1^{(\alpha)})^2$, $\alpha=1,2,3$, we have calculated the first two moments, $m_1^{(\alpha)}$ and $m_2^{(\alpha)}$, from the daughter distribution functions [obtained from equations (\ref{fdistr3ph1})--(\ref{fdistr3ph3})]. From these we have determined, via equation (\ref{distrF}), log-normal distribution functions for three daughter phases (figure~\ref{distr}, green dashed lines). As we seen in figure~\ref{distr},  the daughter distributions are reasonably well approximated by log-normal distributions. For low temperatures, larger differences are observed.

In general, both HTA and MSA give very close predictions for the three-phase behavior of the model at hand.
The only noticeable difference is due to the upper value of the temperature range of the two-phase
equilibrium stability.

\section{Conclusions}
\label{sec:5}

In this paper we have studied the phase behavior of the Yukawa hard-sphere polydisperse mixture at a high
degree of polydispersity using two theoretical methods: HTA and MSA. We confirm the appearance of the
second critical point induced by polydispersity, which was observed earlier using qualitative
van der Waals level of description \cite{Bellier2000,Bellier2001}.
With a further increase of polydispersity, several new
features in the topology of the two-phase diagram emerge: the cloud and shadow curves
intersect twice and each of them forms a closed loop of the ellipsoidal-like shape with liquid and gas
branches of the cloud curve almost coinciding. Upon approaching a certain limiting value of the
polydispersity index $I$ ($I=1.032$), the cloud and shadow curves shrink and disappear.
We have studied the phase behavior of the model beyond this value of polydispersity. At lower values
of the temperature and $I\geqslant 1.032$, polydispersity induces the system to fractionate into three
coexisting phases with different densities: low, intermediate and high. We present and analyze the corresponding phase diagrams together with distribution functions of three coexisting phases.
In general, a good agreement was observed between predictions of the two different theoretical methods, i.e., HTA and MSA. Our results confirm the earlier qualitative predictions for the three-phase coexistence obtained in the
framework of the van der Waals approach  \cite{Bellier2001}.

\clearpage

\appendix

\section{Thermodynamical properties (Helmholtz free energy, chemical
  potential, and pressure) obtained by HTA}

Here, we present the expressions for thermodynamic properties in terms of the moments
(\ref{Mlpol})--(\ref{Mlnmpol}). We have:
\begin{eqnarray}
\label{FRE1b}
\frac{\beta A_{1}}{V}=-2\pi\beta\epsilon_{0}
\frac{Q_{0,A}(z)}{z^3D_0(z)},
\end{eqnarray}
where
\begin{eqnarray}
\label{D0n}
D_0(z)&=&\Delta^2-\frac{2\pi}{z}\left(\Delta+\frac{\pi m_{3,0}}{2}\right)\left(m_{0,\varphi}+\frac{m_{2,0}}{2}\right)\nonumber\\
&&-2\pi \left\{\Delta m_{1,\varphi}+\frac{\pi}{4}\left[m_{2,\varphi}\left(m_{2,0}+2m_{0,\varphi}\right)
-2\left(m_{1,\varphi}\right)^2\right]\right\},\;\;\;\;\;\;\;\;\; \Delta=1-\frac{\pi m_{3,0}}{6},
\end{eqnarray}
\begin{eqnarray}
\label{Q0nm}
Q_{0,A}(z)&=&\bigg\{\Delta\left(m_{0,A}\right)^2+\frac{\pi}{2}
\left[\bigg(m_{3,0}+zm_{2,\varphi}\bigg)
\left(m_{0,A}\right)^2+\left(\frac{z}{2}m_{2,0}
+zm_{0,\varphi}\right)\left(m_{1,A}\right)^2\right]\nonumber\\
&&+z\left(\Delta-\pi m_{1,\varphi}\right)m_{1,A}m_{0,A}\bigg\},
\end{eqnarray}
Differentiating ${\beta A_1}/{V}$ with respect to the density, we get the expression for the chemical potential $\beta\mu_1(\xi)$:
\begin{eqnarray}
\label{chem1b}
\beta \mu_1(\xi)=-\frac{2\pi\beta\epsilon_{0}}{z^3D_0(z)}
\left(\frac{\delta Q_{0,A}(z)}{\delta\rho(\xi)}-\frac{Q_{0,A}(z)}{D_0(z)}
\frac{\delta D_0(z)}{\delta \rho(\xi)}\right),
\end{eqnarray}
where
\begin{eqnarray}
\label{dQ0dro}
\frac{\delta Q_{0,A}(z)}{\delta \rho(\xi)}&=& \bigg\{\pi\left(\frac{m_{3,0}(\xi)}{3}+\frac{zm_{2,\varphi}(\xi)}{2}\right)
\left(m_{0,A}\right)^2+\left(2\Delta+\pi m_{3,0}+\pi zm_{2,\varphi}\right)\nonumber\\
&&\times \; m_{0,A}m_{0,A}(\xi)+\pi z \bigg[\frac{1}{2}\left(\frac{m_{2,0}(\xi)}{2}+m_{0,\varphi}(\xi)\right)\left(m_{1,A}\right)^2
+\left(\frac{m_{2,0}}{2}+m_{0,\varphi}\right)
m_{1,A}m_{1,A}(\xi)\nonumber\\
&&-\left(\frac{m_{3,0}(\xi)}{6}+m_{1,\varphi}(\xi)\right)m_{0,A}m_{1,A}
+\left(\frac{\Delta}{\pi}-m_{1,\varphi}\right)
\left(m_{0,A}m_{1,A}(\xi)+m_{1,A}m_{0,A}(\xi)\right)\bigg]\bigg\},
\end{eqnarray}
\begin{eqnarray}
\label{dD0}
\frac{\delta D_0(z)}{\delta\rho(\xi)}&=&2\pi\bigg\{\frac{1}{3}\pi m_{3,0}(\xi)\left[\frac{1}{2}m_{1,\varphi}-\frac{1}{z}\left(m_{0,\varphi}
+\frac{1}{2}m_{2,0}\right)\right]-
\Delta\left(\frac{1}{6}m_{3,0}(\xi)+m_{1,\varphi}(\xi)\right)\nonumber\\
&&-\left(\frac{1}{2}m_{2,0}(\xi)+m_{0,\varphi}(\xi)\right)
\left[\frac{1}{z}\left(\Delta+\frac{1}{2}\pi m_{3,0}\right)+\frac{1}{2}\pi m_{2,\varphi}\right]-\frac{1}{4}\pi m_{2,\varphi}(\xi)\left(m_{2,0}+2m_{0,\varphi}\right)\nonumber\\
&&+\;\pi m_{1,\varphi}m_{1,\varphi}(\xi)\bigg\}.
\end{eqnarray}

\section{Expressions for thermodynamical properties obtained by MSA}

\begin{eqnarray}
\label{DelN}
\Delta_N=\frac{2\pi\Delta_N[\lambda]}{z^2\Delta(1+\phi^{(1)})+2\pi\Delta_N[\eta]}\;,
\qquad
\Delta_N[\chi]=\chi_{1}\left(\phi_{0}-\frac{z}{2}-\Gamma-\frac{\pi m_{2,0}}{2\Delta}\right)-\chi_{0}(1+\phi_{1}),
\end{eqnarray}
where $\chi_{l}$ stand either for the $\lambda_{l}$ or for the $\eta_{l}$
\begin{align}
\label{phin}
\phi_l&=\rho\int_{0}^{\infty}\rd\xi f(\xi)\sigma^l(\xi)\phi(\xi)\;,&
\phi(\xi)&=\frac{\pi}{2\Delta}\frac{\sigma^2(\xi)\Phi_0(z\sigma)}
{1+\Phi_0(z\sigma)\sigma(\xi)\Gamma},\\
%\end{eqnarray}
%\begin{eqnarray}
\label{etan}
\eta_l&=\rho\int_{0}^{\infty}\rd\xi f(\xi)\sigma^l(\xi)\eta(\xi)\;,&
\eta(\xi)&=\frac{z^2\sigma^3(\xi)\Psi_1(z\sigma)}{1+\Phi_0(z\sigma)\sigma(\xi)\Gamma},\\
%\end{eqnarray}
%begin{eqnarray}
\label{lamn}
\lambda_l&=\rho\int_{0}^{\infty}\rd\xi f(\xi)\sigma^l(\xi)\lambda(\xi)\;, &
\lambda(\xi)&=\frac{A(\xi)}{1+\Phi_0(z\sigma)\sigma(\xi)\Gamma},\\
%\end{eqnarray}
%\begin{eqnarray}
\label{phin1}
m_{l,0}&=\rho\int_{0}^{\infty}\rd\xi f(\xi)\sigma^l(\xi)\;, & \frac{\pi}{6}m_{3,0}&=\eta.
\end{align}
In the above noted expressions, the functions
\begin{eqnarray}
\label{phi_psi}
\Phi_0(x)=\frac{1}{x}(1-\re^{-x})\;, \qquad
\Psi_1(x)=\frac{1}{x^2}\bigg[-1+\bigg(1+\frac{x}{2}\bigg)\Phi_0(x)\bigg]
\end{eqnarray}
were introduced. Functional derivatives are included in the chemical potential:
\begin{eqnarray}
\label{funder1}
\frac{\delta\{\Delta_N\}}{\delta\{f(\xi)\}}&=&\frac{2\pi}{z^2\Delta(1+\phi_{1})+2\pi\Delta_N[\eta]}
\nonumber\\
&&\times\; \left\{\frac{\delta\{\Delta_N[\lambda]\}}{\delta\{f(\xi)\}}-
\Delta_N\left[\frac{\delta\{\Delta_N[\eta]\}}{\delta\{f(\xi)\}}-\left(\frac{\sigma^2(\xi)}{12}-
\frac{\Delta}{2\pi}\phi(\xi)\right)z^2\sigma(\xi)\rho\right]\right\}
\end{eqnarray}
and
\begin{eqnarray}
\label{funder2}
\hspace{-4mm}\frac{1}{\rho}\frac{\delta\{\Delta_N[\chi]\}}{\delta\{f(\xi)\}}\hspace{-2mm}&=&\hspace{-2mm}
\left\{\sigma(\xi)\bigg(\phi_{0}
-\frac{z}{2}-\Gamma-\frac{\pi m_{2,0}}{2\Delta}\bigg)-\phi_{1}-1\right\}\chi(\xi)
\nonumber\\
&&\hspace{-2mm}+\left\{\phi(\xi)+\frac{\pi}{2\Delta}\sigma^2(\xi)\left[\sigma(\xi)
\left(\frac{1}{3}\phi_{0}-
\frac{\pi m_{2,0}}{6\Delta}\right)-1\right]\right\}\chi_{1}-
\left\{\phi(\xi)+\frac{\pi}{6\Delta}\sigma^2(\xi)\phi_{1}\right\}\sigma(\xi)\chi_{0}
\end{eqnarray}
again $\chi$'s stand either for $\{\lambda_{0}, \lambda_{1}, \lambda(\xi)\}$ from equations (\ref{lamn}) or for $\{\eta_{0}, \eta_{1}, \eta(\xi)\}$ from equations (\ref{etan}).

\ukrainianpart

\title{Двофазна та трифазна рівновага в полідисперсній суміші юкавівських твердих сфер.
Високотемпературне та середньосферичне наближення}

\author{Т.В. Гвоздь, Ю.В. Калюжний}
\address{Інститут фізики конденсованих систем НАН України, вул. І.~Свєнціцького, 1, 79011 Львів, Україна}

\makeukrtitle

\begin{abstract}
\tolerance=3000%
Використовуючи високотемпературне (ВТН) та середньосферичне наближення (ССН), досліджено фазову поведінку полідисперсної суміші юкавівських твердих сфер при високих ступенях полідисперсності. Розширено та застосовано схему, яка була розвинена для обчислення фазових діаграм полідисперсних сумішей, які описуються моделями обрізаної вільної енергії, тобто моделями, вільна енергія Гельмгольца яких визначається обмеженим числом узагальнених моментів функції розподілу. В топології двофазних діаграм при високому ступені полідисперсності спостерігаються деякі нові властивості: криві хмари та тіні перетинаються двічі, і кожна з них утворює замкнену петлю еліпсоїдальної форми, причому, рідинна і газова вітки кривої хмари майже збігаються. При певному граничному значенні індекса полідисперсності криві хмари та тіні скорочуються і зникають. Вище, ніж це граничне значення індекса полідисперсності при нижчих температурах зумовлює появу трифазної рівноваги. Представлено і проаналізовано відповідні фазові діаграми разом з функціями розподілу трьох співіснуючих фаз. Спостерігається, загалом, добре узгодження між результатами двох різних теоретичних методів, тобто ВТН і ССН. Наші результати підтверджують якісні передбачення для трифазного співіснування, які були отримані раніше в рамках наближення  ван дер Ваальса.

\keywords полідисперсність, фазове співіснування, колоїдні системи, потенціал Юкави

\end{abstract}

\end{document}